\documentstyle[aps,epsfig]{revtex}
\begin{document}
\draft
\preprint{}
\title{ Self-consistent quantum effects in the quark meson coupling model}
\author{ P.K. Panda$^{\dagger}$\footnote[4]{panda@ift.unesp.br}, 
and F.L. Braghin$^{\ddagger}$\footnote[1]{braghin@axpfep1.if.usp.br}  }
\address{$^\dagger$Instituto de F\'{\i}sica Te\'orica, Universidade Estadual 
Paulista,\\ Rua Pamplona 145, 01405-900 S\~ao Paulo - SP, Brasil}
\address{$^\ddagger$ Instituto de F\'{\i}sica, Universidade de S\~ao Paulo, \\
C.P. 66318 - CEP 05315-970 S\~ao Paulo,  Brasil}
\maketitle
\begin{abstract}
We derive the equation of state of nuclear matter including 
vacuum polarization effects arising from the nucleons and the sigma mesons
in the quark-meson coupling model which incorporates explicitly quark degrees of
freedom with quark coupled to the scalar and vector mesons. 
This leads to a softer equation of state for nuclear matter giving a lower value
of incompressibility than would be reached without quantum effects.
The {\it in-medium} nucleon and sigma meson masses are also calculated in a 
self-consistent manner.
\end{abstract}
\section{Introduction}
Usually the frame of Quantum Hadrodynamics (QHD) \cite{walecka}
is the departure to the study of the nuclear many-body problem
describing nucleons interacting with scalar and vector mesons. This meson
field theory has quite successfully described the properties of the nuclear 
matter and finite nuclei using the mean field approximations for the meson
fields. The vacuum polarization corrections arising from the nucleon fields
as well as the meson fields have also been considered to study the nuclear 
matter. This is one way of obtaining a softer equation of state yielding a lower 
compressibility than would be reached without quantum effects.

While descriptions of the nuclear phenomena have been efficiently formulated
using 
some hadronic degrees of freedom as in QHD, there have been interesting 
observations which reveals the medium modification of the internal structure
of the nucleon. 
For this, degrees of freedom from the fundamental theory of strong interacting
systems, QCD, are expected to be considered. 
Due to the complex structure of this theory we are lead to formulate
effective models which have the main properties and symmetries of QCD
as chiral symmetry and its spontaneously symmetry breaking.  
One of the first models put forward along 
these lines  was the quark-meson coupling (QMC) \cite{qmc,saito} model. It is an 
extension of QHD  which includes the explicit quark
structure of the baryons. This model describes nuclear matter with nucleons 
as non-overlapping MIT bags containing 
quarks inside them which interact with the scalar 
and the vector mesons.  Recently, the model has been extended to finite 
temperatures \cite{panda}, neutron stars \cite{ns} as well as to 
study beyond the mean field level by the incorporation of the exchange 
terms \cite{bkn,ktt}. The density dependent bag constant has also been 
investigated within this frame \cite{bagcon}

In the present work, the vacuum polarization corrections, described in
ref. \cite{glen,mish}, are included in the quark meson coupling model
taking into account the quantum fluctuations of the scalar field 
(as done in \cite{ph4}) self-consistently as well. 
A similar approach is being investigated in the frame of QHD \cite{BCB}. 
The equation of state for dense matter is derived. We organize the paper as 
follows: In section II, we derive the EOS for dense matter including the quantum
effects arising from the nucleons and the $\sigma$-mesons within QMC model.
In section III, we discuss in detail the numerical results obtained in 
the present work and discuss possible outlook.
\section{Theory}
The details of the QMC model have been given in Ref. \cite{qmc,saito,panda}. 
Since we now include the vacuum polarization effects, we give here a few important
steps for  completeness.

In this model, the nucleon in
nuclear matter is assumed to be described by a static MIT bag in which quarks
interact with the scalar ($\sigma$) and the vector ($\omega$) mesons.
The quark field $\psi_q(\vec r,t)$ inside the bag
then satisfies the equation
\begin{equation}
\Big[i\gamma^\mu\partial_\mu-(m_q^0-g_\sigma^q\sigma)-g_\omega^q\omega\gamma^0
\Big]\psi_q(\vec r,t)=0,
\end{equation}
where $m_q^0$ is the current quark mass and $g_\sigma^q$ and $g_\omega^q$
are the quark couplings with the $\sigma$ and $\omega$ mesons.
 
The normalized ground state for a quark (in an s-state) in the bag,
which has a radius $R$, is given as
\begin{equation}
\psi_q(\vec r,t)=N \exp(-i\frac{\epsilon_q t}{R})
\left(
\begin{array}{c} j_0(xr/R)
\\ i\beta_q \vec \sigma \cdot \hat r j_1(xr/R)
\end{array}
\right) \frac{\chi_q}{\sqrt {4\pi}},
\label{wavefn}
\end{equation}
where $x$ is the dimensionless quark momentum and 
the single particle quark energy, in units of $R^{-1}$, is
\begin{equation}
\epsilon_q=\Omega_q+g_\omega^q\omega R, \quad
\beta_q=\sqrt{\frac{\Omega_q-R m_q^*}{\Omega_q+R m_q^*}},
\end{equation}
with
$\Omega_q=(x^2+R^2{m^*_q}^2)^{1/2}$;
$m^*_q=m^0_q-g_\sigma^q\sigma$ is the effective quark mass,
$\chi_q$ is the quark spinor and $N$ is the
normalization factor.
 
The boundary condition at the bag surface is given by:
\begin{equation}
i\gamma\cdot n \psi_q=\psi_q.
\end{equation}
This, for the ground state, reduces to
\begin{equation}
j_0(x)=\beta_q j_1(x)
\label{x}
\end{equation}
which determines the eigen frequency, $x$, of this lowest mode in the medium.
The form of the quark wavefunction in equation (\ref{wavefn}) is almost 
identical to that of the solution in free space. However the parameters
in the expression have been substantially modified by the surrounding nuclear
medium. Thus the quarks in the nucleon embedded in the nuclear medium are
more relativistic than those in a free nucleon.

The energy of the nucleon bag is
\begin{equation}
M^*=3\frac{\Omega_q}{R}-\frac{Z}{R}+\frac{4}{3}\pi R^3 B,
\label{ebag0}
\end{equation}
where $B$ is the bag constant and $Z$ parametrizes the sum of the center-of-mass
(c.m.) motion and the gluonic corrections. Note that this center-of-mass treatment
is different from that of Jin and Jennings \cite{bagcon}.
The bag radius $R$ is that which minimizes the nucleon bag energy 
through
\begin{equation}
\frac{\partial M^*}{\partial R}=0.
\label{rr}
\end{equation}

We now proceed to study the equation of state (EOS) for nuclear
matter including the vacuum polarisation effects from nucleon and sigma mesons
at zero temperature. The details of the theory have already been discussed 
in ref. \cite{mish}. Only a few important steps are given here.
The energy density after subtracting out the pure vacuum
contribution then becomes
\begin{equation}
\epsilon_0 = \epsilon_{MFT}+\Delta \epsilon,
\label{enh}
\end{equation}
with
\begin{equation}
\epsilon_{MFT}=\frac{g}{(2\pi)^3}\int\limits_{|{\bf k}|<k_F} 
d {\bf k}(k^2+{M^*}^2)^{1/2} +\frac{1}{2}m_\sigma^2 \sigma_0^2
+\frac{1}{2}m_\omega^2 \omega_0^2 ,
\end{equation}
and
\begin{equation}
\Delta \epsilon=-\frac{g}{(2\pi)^3}\int d {\bf k} \bigg[(k^2+{M^*}^2)
^{1/2}- (k^2+M^2)^{1/2}-\frac{g_{\sigma }~\sigma_0 M}{(k^2+M^2)^{1/2}}
\bigg].  \label{div}
\end{equation}
The above expression for the energy density is divergent.
After renormalisation  by adding the counter terms \cite{chin},
we have the expression for the finite renormalised energy density,
\begin{equation}
\epsilon_{ren}= \epsilon_{MFT}+\Delta \epsilon_{ren},
\end{equation}
where
\begin{equation}
\Delta \epsilon_{ren}= -\frac{g}{16 \pi^2}\bigg[
{M^*}^4 \ln \bigl(\frac{M^{*}}{M}\bigr)+M^3 (M-{M^*})
-\frac{7}{2}M^2  (M-{M^*})^{2}
+ \frac{13}{3} M  (M-{M^*})^3 -\frac{25}{12} M  (M-{M^*})^4\bigg].
\label{rhf}
\end{equation}
The baryon density is given by
\begin{equation}
\rho_B=\frac{g k_F^3}{6\pi^2}
\end{equation}
In the above, $g$  is the spin-isospin degeneracy factor 
which is equal to $4$ for nuclear matter and to $2$ for neutron matter.

Next, we consider the quantum corrections due to the scalar mesons. 
Including a quartic scalar self-interaction, the Hamiltonian density
for the scalar mesons becomes 
\begin{equation}
{\cal H}_\sigma= \frac{1}{2}\partial_\mu \sigma \partial^\mu \sigma
+\frac{1}{2} m_\sigma ^2 \sigma^2+\lambda \sigma^4,
\label{lwsgm}
\end{equation}
with $m_\sigma$ and $\lambda$ being the bare mass and coupling 
constant respectively. 
We calculate the expectation value of the Hamiltonian density and perform
the renormalisation according to prescription of ref. \cite{politzer}.
The resulting gap equation for $M_\sigma^2$, which minimizes the energy, 
in terms of the renormalised parameters 
$m_R^2$ and $\lambda_R$ can be written as
\begin{equation}
M_\sigma^2=m_R^2+12\lambda_R\sigma_0^2+12\lambda_R I_f(M_\sigma),
\label{mm2}
\end{equation}
where
\begin{equation}
I_f(M_\sigma)=\frac{M_\sigma^2}{16\pi^2}\ln \Big(\frac{M_\sigma^2}{\mu^2} \Big).
\label{if}
\end{equation}
Using the above equations  we obtain the
energy density for the $\sigma$ in terms of $\sigma_0$ which is given by
\begin{equation}
\epsilon_\sigma=3\lambda_R\Big(\sigma_0^2+\frac{m_R^2}{12\lambda_R}\Big)^2
+\frac {M_\sigma^4}{64\pi^2}\Biggl(\ln\Big(\frac{M_\sigma^2}{\mu^2}\Big)
-\frac{1}{2} \Biggr)
-3\lambda_R I_f^2-2\lambda\sigma_0^4.
\label{vph}
\end{equation}
Where $\mu$ is a mass scale. The above expression is given in terms of
$\sigma$ mass $m_R$ and  $\lambda_R$ except for the
last term which is still in terms of the bare coupling constant $\lambda$
and did not get renormalised because of the structure of the gap
equation \cite{pi}.  However, from the renormalisation procedure
it is easy to see in this last work that when $\lambda_R$ is kept fixed, 
the bare coupling $\lambda \rightarrow 0_-$.
Therefore the last term in eq. (\ref{vph}) will be neglected
in the numerical calculations. 

After subtracting the vacuum contribution we obtain:
\begin{eqnarray}
\Delta \epsilon_\sigma &=& \epsilon_\sigma-\epsilon_\sigma(\sigma_0=0)
\nonumber \\
&=& \frac{1}{2} m_R^2 \sigma_0^2+ 3\lambda_R \sigma_0^4 
+\frac {M_\sigma^4}{64\pi^2}
\Biggl(\ln\Big(\frac{M_\sigma^2}{\mu^2}\Big)-\frac{1}{2} \Biggr)
-3\lambda_R I_f^2 -\frac {M^4_{\sigma,0}}{64\pi^2}
\Biggl(\ln\Big(\frac{M_{\sigma,0}^2}{\mu^2}\Big)-\frac{1}{2} \Biggr)
+3\lambda_R I_{f0}^2,
\label{vph0}
\end{eqnarray}
where $M_{\sigma,0}$ and $I_{f0}$ are the expressions
given by eqs. (\ref{mm2}) and (\ref{if}) with $\sigma_0=0$.

The energy density and pressure 
with baryon and the sigma condensate $\sigma_0$ are  respectively done by
\begin{equation}
\epsilon_{ren}=\epsilon_0^ {finite}+\Delta\epsilon_{ren},
\end{equation}
and
\begin{equation}
P=\frac{g}{3 (2\pi)^{3}}\int\limits_{|{\bf k}|<k_F} 
d{\bf k} \frac{k^2}{(k^2+{M^*}^2)^{1/2}} + 
\frac{1}{2}m_\omega^2 \omega_0^2 -\Delta\epsilon_\sigma
-\Delta\epsilon_{ren},
\end{equation}
where
\begin{equation}
\epsilon_0 ^{finite}
=\frac{g}{(2\pi)^3}\int\limits_{|{\bf k}|<k_F} 
d{\bf k} (k^2+{M^*}^2)^{1/2} 
+\frac{1}{2}m_\omega^2 \omega_0^2 +\Delta\epsilon_\sigma
\end{equation}
with $\Delta\epsilon_{ren}$ given by eq. (\ref{rhf}) and 
$\Delta\epsilon_{\sigma}$ by eq. (\ref{vph0}).

The energy density from the $\sigma$ field as given by eq.
(\ref{vph0}) is still in terms of the renormalisation scale $\mu$
which is arbitrary. We choose this to be equal to 
the renormalised sigma mass $m_R$ in doing the numerical
calculations. This is because changing $\mu$ would mean changing 
the quartic coupling $\lambda_R$, and $\lambda_R$ here
enters as a parameter to be chosen to give the incompressibility for 
nuclear matter in the correct range.
The  parameters $g_\sigma^q$ and $g_{\omega}$ are 
fitted so as to describe the ground-state properties of nuclear matter
correctly. For a given baryon density $\rho_B$, the energy density,
the density dependent radius of the nucleon and the nucleon effective mass are
calculated at zero temperature. 
\section{Results and discussion}
We now proceed with the numerical calculations for the nuclear matter.
We start fixing the bag properties in the vacuum. They are given 
in Table 1. We next calculate the ground state properties of the nuclear 
matter and fit the scalar and vector coupling constants $g_\sigma^q$ and 
$g_\omega (= 3g_\omega^q)$ to get the correct saturation properties for a
given renormalised $\sigma$ mass and coupling, $m_R$ and $\lambda_R$. The 
omega and sigma couplings for given $\lambda_R$ are tabulated in Table II.
Results from Relativistic Hartree approximation (RHA) are also shown.
Little change is noted in the values of the parameters for the range 
of $\lambda_R$ which is considered here.

Using these values, we plot the binding energy $(E_B=\epsilon/\rho_B-M_N)$
for nuclear matter as a function of density in Figure 1. 
In the same figure we also plot the results for the relativistic Hartree 
approximation (RHA). Clearly, including
baryon and $\sigma$-meson quantum corrections leads to a softer equation of 
state which is further softer for a higher value of $\lambda_R$. 
The equation of state, pressure, $P$ versus as a function of energy density
$\epsilon$ is displayed in Fig. 2 for the different cases. For comparison, the
causal limit $P=\epsilon$ is also shown in the figure. All the cases studied
here respect the causal condition $\partial P/\partial\epsilon \le 1$, so
that the speed of sound remains lower than the speed of light.

In figure 3 we plot the effective nucleon mass as a function of density.
At the saturation density we get $M^*=0.817 M$ and $0.83 M$ for $\lambda_R=3.0$
and $\lambda_R=4.5$ respectively. These values may be compared with the 
results of $M^*=0.775 M$ for normal QMC model and of $0.793 M$ with the 
relativistic Hartree approximation in QMC model. This influence is much higher 
at high nuclear densities. We can conclude that quantum effects, at the level 
we consider, increase the effective mass $M^*$.

We plot the {\it in medium} 
effective radius of the nucleon ($R^*$) as a function of density
in figure 4. $R^*$ is also increased with relation to the mean field 
approximation mainly for higher densities. 
It is possible to understand this result
as an effect which prevents QMC from deconfinement at high densities.

In figure 5, we plot the {\it in-medium} 
$\sigma$-meson mass, $M_\sigma$, as a function
of density. $M_\sigma$ increases with density as $\lambda_R$ is positive.
The higher is the coupling, higher is 
the increase of the sigma mass with density.
This would be rather an indication of a further chiral symmetry breaking
instead of its restoration \cite{FLB01c}.

To summarize, we have used a non-perturbative approach to include the quantum
effects in nuclear matter in the framework of QMC. The calculations of the
scalar meson quantum corrections was done here in a self-consistent manner
including  multi-loop effects. This leads to a softening of the equation of
state. We have also calculated the effective mass of the $\sigma$-meson as
modified by the quantum corrections. The effective sigma mass is seen to
increase with density. The results may be suggesting that, at high densities, 
quantum fluctuations prevents the model from the chiral symmetry restoration
as well as from deconfinement.
These results will be extensively studied in another work where the coupling 
constant $\lambda_R$ and its influence on the chiral symmetry behavior -whose
order parameter can be considered to be $\sigma_0$-  at high
densities and temperatures will be shown in the frame of the model 
worked out here. 
\section{Acknowledgments}
P.K.P. would like to acknowledge FAPESP (Processo- 99/08544-0)
for financial support and the IFT, S\~ao Paulo, for kind hospitality.
F.L.B.  acknowledges support from FAPESP (Processo- 97/01317-3).
\begin{table}
\caption{Parameters used in the calculation.}
\begin{tabular}{ccccccc}\hline
M (MeV)&$m_q$ (MeV) & $R$ (fm) & $B^{1/4}$ (MeV) & $z_0$ & $m_R$ (MeV) 
& $m_\omega$ (MeV) \\ \hline
939. &0 & 0.6 & 211.3 & 3.987 & 550 & 783 \\ \hline
\end{tabular} 
\end{table}

\begin{table}
\caption{quark-sigma, omega-nucleon couplings are used for 
different cases in our calculation. Effective nucleon mass, effective radius
are given for different sets.\vspace{0.5in}}
\begin{tabular}{ccccccc}
\hline
~case~ &~$g_s^q$~ &~ $g_\omega$~&~$M^*/M_N$~&~$R^*$ (fm)~&~$K$(MEV) \\
\hline
normal QMC&5.98575&8.96259&0.775&0.5961& 290.9  \\
RHA&5.77097&8.3935&0.793&0.5967& 272.1 \\
$\lambda_R=3.0$&5.51202&7.76908&0.817&0.5975& 256.1  \\
$\lambda_R=4.5$&5.37293&7.36239&0.830&0.5978& 244.8 \\
\hline
\end{tabular}                
\end{table}
\begin{figure}[htb]
\epsfxsize=0.8\textwidth
\begin{center}
\epsfbox{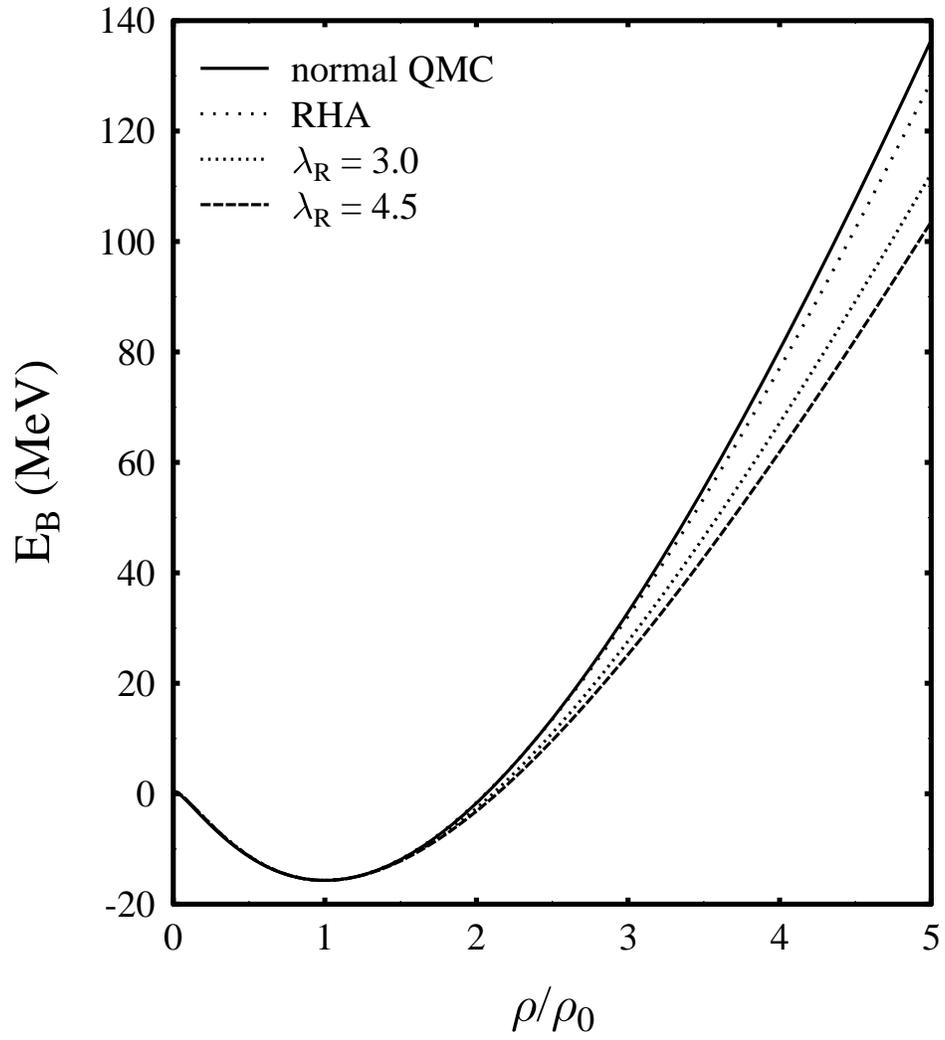}
\end{center}
\caption{The binding energy of the nuclear matter as a function of densities.
Including the quantum corrections  give a softer equation of state.}
\end{figure}
\begin{figure}[htb]
\epsfxsize=0.8\textwidth
\begin{center}
\epsfbox{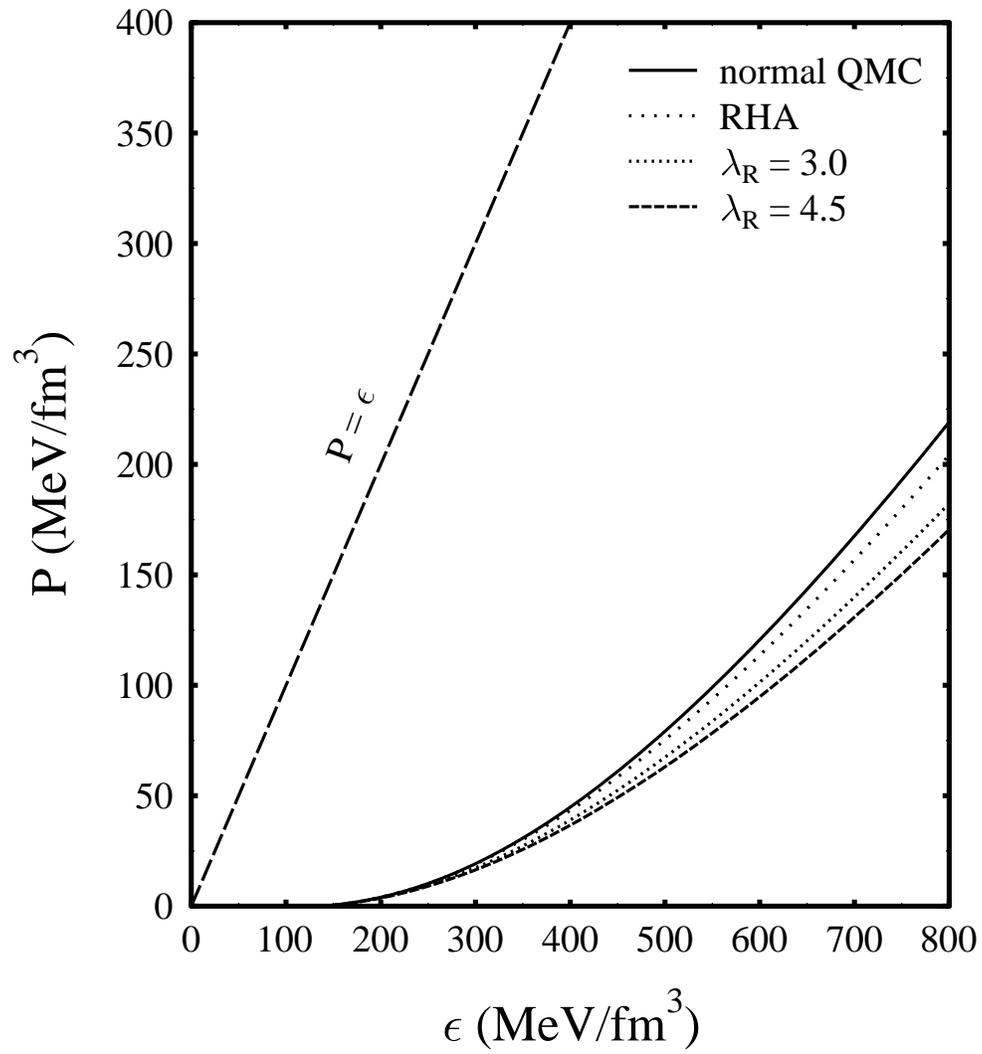}
\end{center}
\caption{The pressure versus energy density of the nuclear matter.}
\end{figure}
\begin{figure}[htb]
\epsfxsize=0.8\textwidth
\begin{center}
\epsfbox{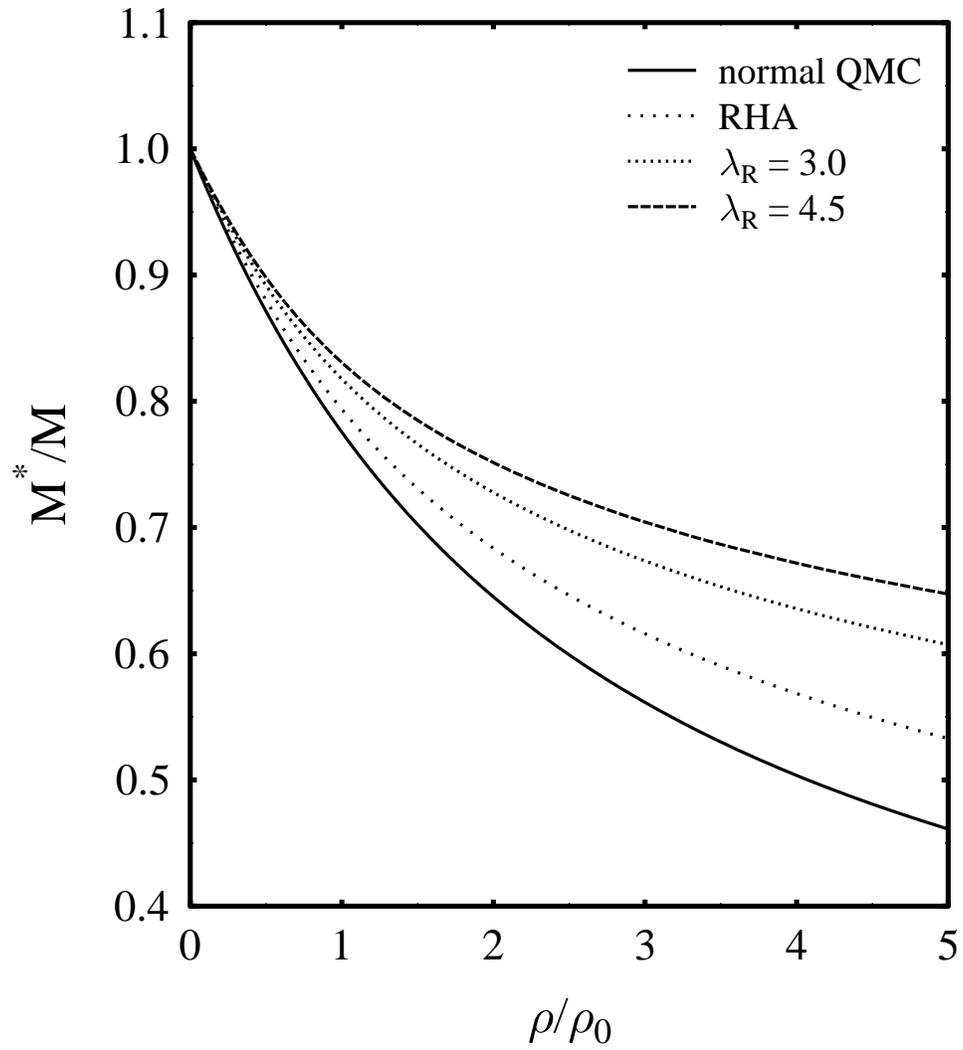}
\end{center}
\caption{Effective baryon masses in the medium.}
\end{figure}
\begin{figure}[htb]
\epsfxsize=0.8\textwidth
\begin{center}
\epsfbox{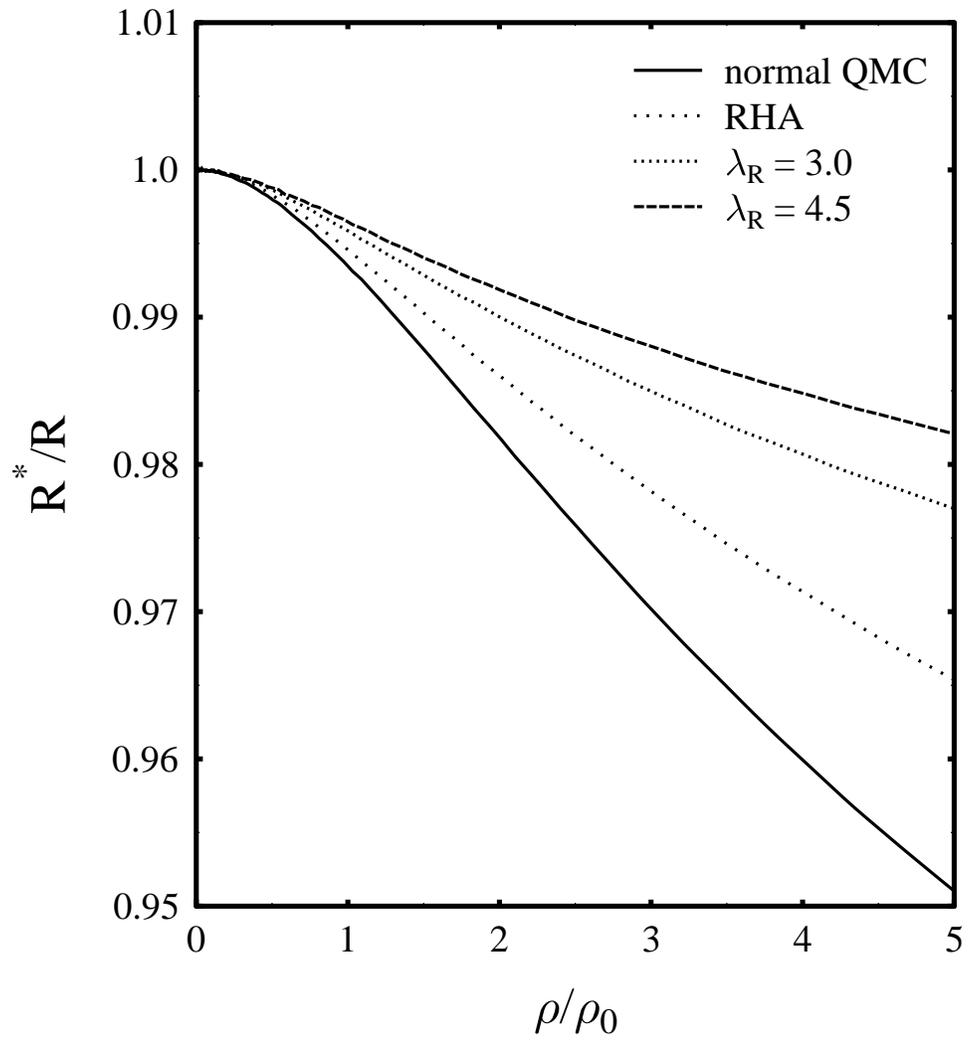}
\end{center}
\caption{The effective radius of the nucleon as a function of densities.}
\end{figure}
\begin{figure}[htb]
\epsfxsize=0.8\textwidth
\begin{center}
\epsfbox{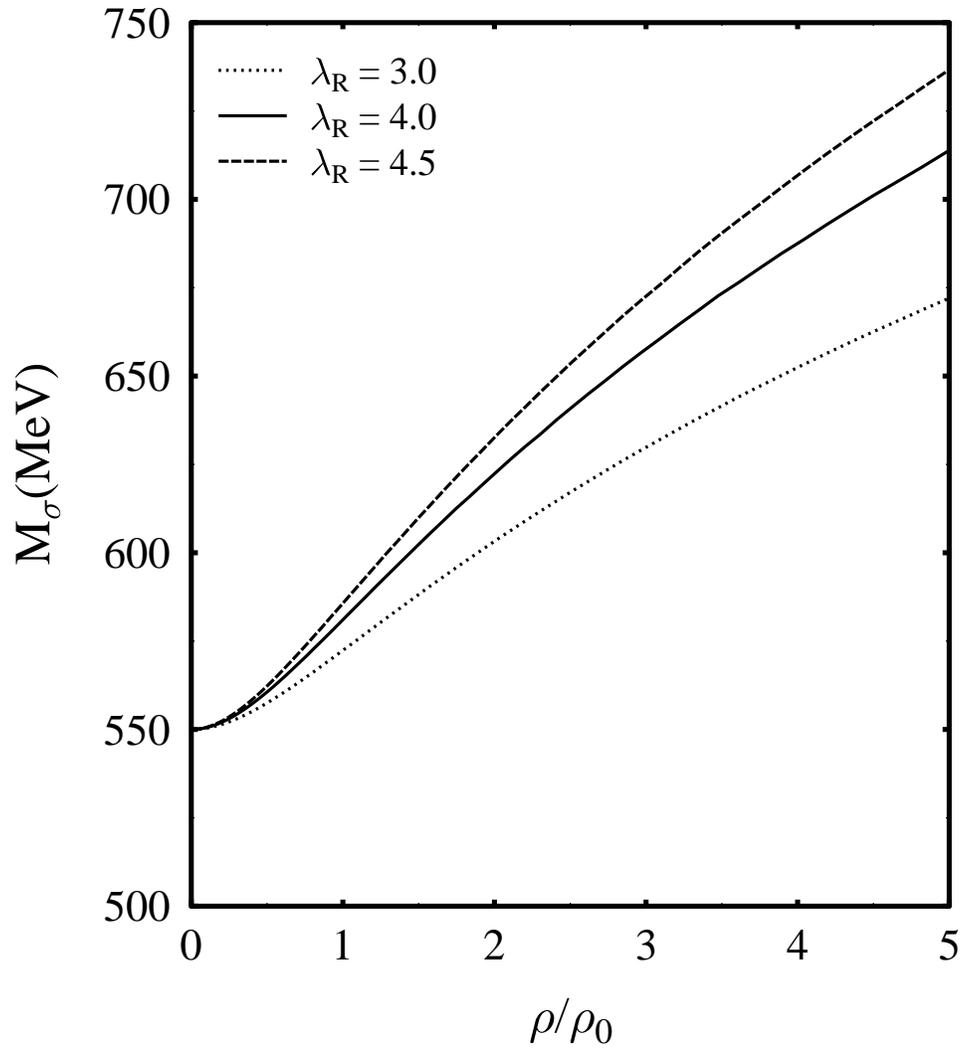}
\end{center}
\caption{In medium scalar meson mass versus baryon density.}
\end{figure}
\end{document}